\documentclass[footinbib,twocolumn,showpacs,amsmath,amstex,amssymb,mathfonts,superscriptaddress,prx]{revtex4}
\usepackage{graphicx}
\usepackage{color}
\usepackage{bm}

\usepackage{graphicx}
\usepackage{color}
\usepackage{bm}
\usepackage{amsmath}
\usepackage{amssymb}
\usepackage{amsthm}
\usepackage{amsfonts}
\usepackage{multirow}
\usepackage{makecell}
\usepackage{float}
\usepackage{comment}
\usepackage{enumitem}
\usepackage{verbatim}
\usepackage{mathtools}

\begin{document}

\title{Protected edge modes based on the bulk and boundary renormalization group: A relationship between duality and generalized symmetry}
\author{Yoshiki Fukusumi}
\affiliation{Division of Physics and Applied Physics, Nanyang Technological University, Singapore 637371.}

\date{\today}
\begin{abstract}

We propose a theoretical formulation of protected edge modes in the language of quantum field theories based on the contemporary understanding of the renormalization group. We use bulk and boundary renormalization arguments which have never captured enough attention in condensed matter physics and related fields. We revisit various exotic bulk and boundary phenomena in contemporary physics, and one can see  the conciseness of our formulations. Moreover, in the systems with open boundaries in general space-time dimensions, we also analyze their implications under general duality implemented by the shift of defects corresponding to generalized symmetries, including higher-form, non-invertible symmetries, in principle. Our formulation opens up a new paradigm to explore the systems with protected edge modes in the established language of the renormalization group.
\end{abstract}

\maketitle

\section{Introduction}

Edge modes of topologically ordered systems are one of the most fundamental and exotic objects in contemporary theoretical physics (and part of experimental physics), including condensed matter, high energy, and mathematical physics. Typically, one can relate them to exotic particles, like Majorana fermion, parafermion (and non-abelian anyon in general)\cite{Kitaev:2001kla,Fendley:2012vv}. Historically, one can see the appearance of such ``emergent" boundary degrees of freedom at the edges in the Heisenberg and AKLT chain\cite{Affleck:1987vf,TKennedy_1990,White:1993zz}, correspondence between topological quantum field theory (TQFT) and conformal field theory\cite{Witten:1988hf} and the edges of quantum Hall systems\cite{Moore:1991ks}. The related appearance of such boundary degrees of freedom from entanglement cuts, typically in symmetry-protected topological (SPT) phases, has captured the attention of condensed matter physicists\cite{Pollmann_2010}. It may be worth noting that these ``emergent" particles have captured the attention of both theorists and experimentalists as a building block of topological quantum computation\cite{Mong_2014}. 

More recently,  strong zero modes \cite{Fendley_2016} which may be realized in a nonequilibrium setting have been proposed and studied \cite{Tan:2022vaz,Mitra:2023xdo,Samanta:2023fvs}. One can understand them based on the earlier works of topological defects that satisfy the non-abelian fusion rule (which can be interpreted as defects analog of non-invertible symmetries in some cases)\cite{Petkova:2000ip}. We also note the theoretical work \cite{Graham:2003nc} as a pioneer work in this research direction (but from a different motivation), whereas it has been rarely mentioned as such in the literature.

However, the theoretical (or mathematical, to some extent) formulation of such ``emergent" particles at boundary, or protected edge modes is still far from established. Moreover, understanding of them based on the bulk and boundary (or defect) renormalization group (RG)\cite{Dorey:1999cj,Dorey:2004xk,Dorey:2009vg,2009JPhA...42W5403F,Chang:2018iay}, which can be the starting point to formulate the phenomena, has been studied only recently and partially\cite{2013PhRvX...3b1009L,Lichtman:2020nuw,Fukusumi:2020irh,Moradi:2022lqp,Fukusumi:2022xxe,Fukusumi_2022,Fukusumi_2022_c} (whereas the non-protected edge modes, called gapped edge modes have been studied widely\cite{Wang:2012am,Lan:2014uaa,Kaidi:2021gbs}). Related works with emphasis on categorical descriptions can be seen in \cite{Kong:2019byq,Kong:2019cuu,Chatterjee:2022tyg}.

In this work, we propose a field theoretic formulation of protected edge modes based on the bulk and boundary RG which can be understood as a quantum field theory containing both bulk and boundary perturbations. Our formulation clarifies the ``emergent" symmetry or degree of freedom in a quantitive way, and its relation to bulk and boundary RG which can be evaluated by finite size scaling. Hence, based on our formalism, such ``emergent" phenomena can be evaluated by numerical calculations, such as truncated conformal space approach \cite{Yurov:1989yu,Yurov:1991my,Hogervorst:2014rta,James_2018} in principle. Our work may serve as a building block to formulate protected edge modes rigorously. We expect that our formalism gives a useful theoretical framework applicable to the analysis of lattice models and for future experimental realizations.

The rest of the manuscript is organized as follows. In Sec.\ref{section_RG}, we formulate the protected edge modes by using the bulk and boundary RG. We concentrate on $1+1$ dimensional systems, but the discussion can be easily generalized to general space-time dimensions. We analyze various phenomena related to protected edge modes in contemporary condensed matter theory in a concise way. In Sec.\ref{section_duality}, we revisit the theoretical formulation of dualities in a $1+1$ dimensional system. One can identify the topological defect at criticality as the generator of duality in the system with open boundaries. In Sec.\ref{section_higher_dimension}, we generalize the arguments in previous sections to higher space-time dimensional systems. We also propose dualities in higher dimensional QFTs and their implications. Finally, in Sec.\ref{section_conclusion}, we make a concluding remark with open problems in the fields.

\section{Bulk and boundary renormalization group theoretic formulation of protected edge modes}
\label{section_RG}
In this section, we restrict our attention to a $1+1$ dimensional system with left and right boundaries to avoid complications. However, it should be stressed that one can straightforwardly generalize our argument to higher dimensional models as we will discuss in Sec.\ref{section_higher_dimension}. For convenience, we label the left and right boundary as lower indices ``$0$" and ``$1$" respectively, and the time (or inverse temperature) as periodic to concentrate on the Hamiltonian formalism of quantum field theory\cite{Cardy:1986ie}.
Let us introduce the following class of models,

\begin{align}
&\mathcal{H}_{\alpha_{0}, \alpha_{1}}, \\
&H_{\text{CFT},\alpha_{0}, \alpha_{1}}+H_{\text{B}}+H_{\text{b},0}+H_{\text{b},1}
\end{align}
where $\mathcal{H}_{\alpha_{0},\alpha_{1}}$ determined by the boundary conditions is the Hilbert of the model without perturbations, and $H_{\text{CFT},\alpha_{0}, \alpha_{1}}$ is the Hamiltonian of a conformal field theory (CFT) with the left (right) boundary condition labeled by index $\alpha_{0}$ ($\alpha_{1}$), and $H_{\text{B}}, H_{\text{b},0} ( H_{\text{b},0})$ are bulk and left (right) boundary perturbation terms respectively. In the contemporary condensed matter theory, it has been noticed that the form of $\alpha$ can be outside of Cardy's states which describe a class of stable boundary conditions\cite{Verresen:2019igf,Runkel:2020zgg,Fukusumi:2020irh,Fukusumi:2021zme,Weizmann,Smith:2021luc,Fukusumi_2022_c}. For the readers interested in the structure of the Hilbert space $\mathcal{H}_{\alpha_{0},\alpha_{1}}$, we note \cite{Cardy:2004hm,Petkova:2000dv} as reviews and \cite{DiFrancesco:1997nk} as a textbook. More strictly speaking, there exist open problems about the definition of the boundary operators and their realization in a lattice model. For example, the appearance of logarithmic operators even in the unitary models has been reported \cite{Gori:2018gqx,Gori:2017cyq}. However, to proceed with the discussions, let us assume that we have already obtained a class of treatable models.

One can implement a series of non-Cardy states by applying topological defects to Cardy states recursively, and the author and the collaborator called them Graham-Watts (GW) states\cite{Fukusumi:2020irh} which have first appeared in \cite{Graham:2003nc}. Let us assume the existence of boundary RG flow between Cardy states $|\alpha\rangle$ triggered by $\phi_{a}$ where $a$ is a label of primary fields. The most significant point of GW states is that this RG can be mapped to that of GW states $D_{b}|\alpha\rangle$ triggered by the field $D_{b}\phi_{a}$ where $D_{b}$ is the topological defect labeled by the primary field $\phi_{b}$\cite{Graham:2003nc,Kojita:2016jwe}. Hence the boundary RG analysis of Cardy states \cite{Affleck:1990by,Recknagel:2000ri,Graham:2000si,Fredenhagen:2002qn,Fredenhagen:2003xf} result in the boundary RG analysis of GW states.

Recently, several proposals to formulate RG based on the RG domain walls\cite{CRNKOVIC1990637,Gaiotto:2012np} have been proposed\cite{Klos:2019axh,Klos:2021gab,Kikuchi:2022rco,Zeev:2022cnv,Poghosyan:2023brb}. We also note several works studying RG domain wall \cite{Poghosyan:2014jia,Stanishkov:2016pvi,Stanishkov:2016rgv,Poghosyan:2022mfw,Poghosyan:2022ecv}, its BCFT analog\cite{Konechny:2012wm,Konechny:2019wff,Konechny:2020jym}, and related earier works on conformal interface and multijunction problems\cite{2007JHEP...04..095Q,2014NuPhB.885..266K,2015JHEP...07..072K}.
Based on these research directions, one can formally understand the RG as the following mapping of UV QFT, $\text{QFT}_{UV}$, and a general quantum field theory $\text{QFT}_{1}$ labeled by the upper index $(1)$,
\begin{align}
&\mathcal{H}_{\text{QFT}_{UV},\alpha_{0},\alpha_{1}}\rightarrow \mathcal{H}_{\text{QFT}_{1},\alpha_{0}^{(1)},\alpha_{1}^{(1)}} \\
\begin{split}
&H_{\text{QFT}_{UV},\alpha_{0},\alpha_{1}}+H_{\text{B}_{UV}}+H_{\text{b}_{UV},0} +H_{\text{b}_{UV},1}\\
&\rightarrow H_{\text{QFT}_{1},\alpha^{(1)}_{0},\alpha_{1}^{(1)}}+H^{(1)}_{\text{B}}+H^{(1)}_{\text{b},0} +H^{(1)}_{\text{b},1}
\end{split}
\end{align}
where $\mathcal {H}$ is the Hilbert space of each theory specified by the boundary conditions, and we have assumed general boundary conditions. Rigorous construction of such boundary conditions and states contains difficulty (even when restricting our attention to CFT, there can exist a series of nontrivial boundary states, called symmetry breaking boundary conditions\cite{Quella:2002ns,Quella:2002fk,2002JHEP...06..028Q,Blakeley:2007gu}), so let us assume the existence of them again. After each RG procedure, perturbations tend to become irrelevant, and finally, all bulk and boundary perturbations become irrelevant (For simplicity let us avoid the case with dangerous irrelevant perturbations \cite{2018JHEP...10..108G,PhysRevB.99.195130}). We label this theory in which all perturbation becomes irrelevant as IR QFT and label them by using ' in this work. We restrict our attention to the following flow of the UV CFT to the IR QFT,
\begin{align}
&\mathcal{H}_{\text{CFT}_{UV},\alpha_{0},\alpha_{1}}\rightarrow \mathcal{H}_{\text{QFT}_{IR},\alpha'_{0},\alpha'_{1}}, \\
\begin{split}
&H_{\text{CFT}_{UV},\alpha_{0},\alpha_{1}}+H_{\text{B}_{UV}}+H_{\text{b}_{UV},0} +H_{\text{b}_{UV},1}\\
&\rightarrow H_{\text{QFT}_{IR},\alpha'_{0},\alpha'_{1}}+H'_{\text{B}}+H'_{\text{b},0} +H'_{\text{b},1}.
\end{split}
\end{align}
(However, the discussion can be applied to RG started from general QFTs.) When restricting our attention to the boundaries, one can see the following bulk and boundary RG,
\begin{equation}
|\alpha_{i}\rangle \rightarrow|\alpha'_{i}\rangle, \ i=0,1.
\end{equation}
If it is not necessary, we drop off the lower index $i$ in the discussion of boundary RG.

As a concrete formulation, it is necessary to consider Hilbert space sufficiently large by choosing the boundary condition $\alpha_{i}$, corresponding to the situation on specific models, typically on the lattice models. For this purpose, it is inevitable to start the RG argument from non-Cardy states in general. We assume the following conditions,
\begin{itemize}
\item{ The UV CFT has a sufficiently large central charge.}
\item{ The UV boundary condition has sufficiently large boundary entropy, called $g$-value \cite{Affleck:1991tk}, such that the $g$-value decreases monotonically under bulk and boundary RG.}
\end{itemize}
The first point is a usual condition to consider RG's argument \cite{Zamolodchikov:1986gt} (when considering its proof, things can become complicated \cite{Barnes:2004jj,Friedan:2009ik,Gukov:2015qea}, so let us assume this). For example, when considering spin $S$ $SU(2)$ chain, by analyzing its mapping to $\{ SU(2)_{1}\}^{2S}(= \{U(1)\}^{2S})$ Wess-Zumino-Witten (WZW) model\cite{PhysRevB.34.6372,Schulz}, one can see the $Z_{2}$ anomaly and the corresponding Lieb-Schultz-Mattis type anomaly. One can apply this observation to $SU(N)$ Haldane conjecture for example\cite{Haldane:1982rj,Haldane:1983ru,Lecheminant:2015iga,Lajko:2017wif,Yao:2018kel,Wamer:2019oge,Kikuchi:2022ipr,Fukusumi_2022_c}.

\begin{figure}[htbp]
\begin{center}
\includegraphics[width=0.5\textwidth]{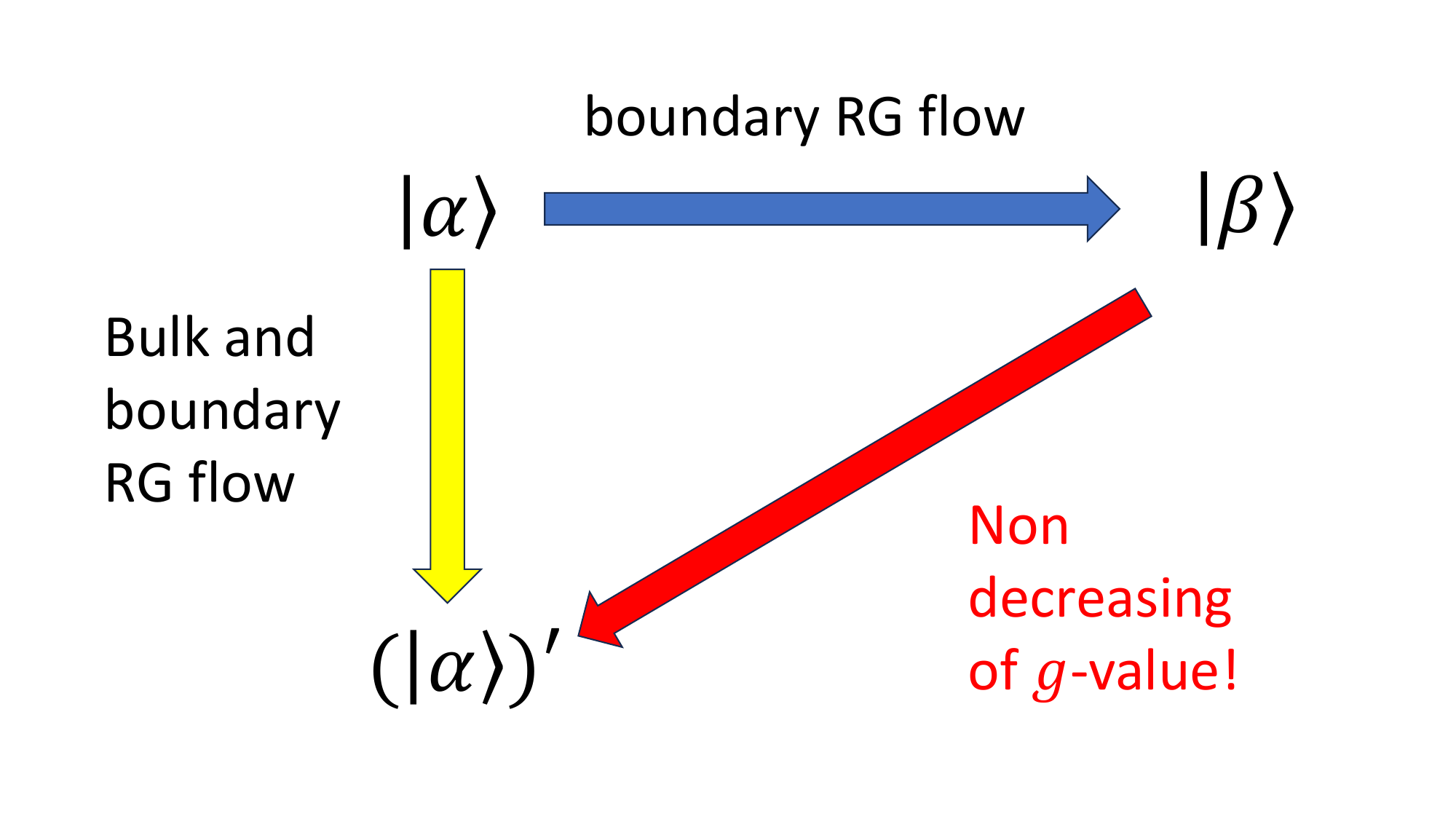}
\caption{Diagramatic picture of emergent edge degree of freedom. $|\alpha\rangle$, $|\beta\rangle$ corresponds to boundary states of the UV CFT and $(|\alpha\rangle)'$ corresponds to that of IR theory, and each arrow corresponds to the RG flows. For the RG flows represented by the yellow and blue arrows, the monotonic decreasing of the degrees of freedom occurs ($c$-theorem and $g$-theorem). However, for the red arrow, the boundary $g$-value can show nondecreasing, because the relevant boundary operators that trigger the boundary flow $|\alpha\rangle\rightarrow |\beta\rangle$ can become irrelevant.}
\label{bulk_boundary}
\end{center}
\end{figure}

The second point may be more unfamiliar to the readers. The $c$-theorem \cite{Zamolodchikov:1986gt} is a guiding principle to consider a bulk RG flow. This indicates a monotonic decrease in the bulk degree of freedom. One can consider the boundary analog of RG, i.e. $g$-theorem\cite{Affleck:1991tk,Friedan:2003yc,Nakayama:2012ed,Casini:2022bsu}. The $g$-theorem indicates that the boundary degree of freedom should decrease under boundary RG, \emph{under fixing bulk universality class}. In other words, without this fixing, the $g$-value can show nondecreasing or even increasing \cite{Green:2007wr}. (In non-Hermitian systems corresponding to nonunitary or complex CFTs, breaking of $g$-theorem has been proposed\cite{Nakagawa:2018ioz,Han:2023ygh}. However, these works have been lacking attention to this fixing of bulk universality class. Moreover, unfortunately, they have never paid attention to the existing literature.) This can be considered as a characteristic property of protected edge modes, or ``emergent" (generalized) symmetry at the boundary, but this can complicate the bulk and boundary RG analysis. Hence, we formulate the protected edge mode by taking both bulk and boundary degree of freedom sufficiently large and by observing this emergence of boundary degree of freedom as stabilization of unstable UV boundary conditions. This stabilization can be formulated by the transformation of boundary-relevant perturbation in UV CFT to the irrelevant ones in IR under the bulk and boundary RG (FIG.\ref{bulk_boundary}). This picture can be thought of as a modern view of protected edge modes\cite{Fukusumi:2020irh,Fukusumi_2022_c}.

There exist several works studying bulk and boundary RG flow or related RG flow of defects\cite{Dorey:1999cj,Dorey:2004xk,Dorey:2009vg,2009JPhA...42W5403F,Chang:2018iay}. Their main concern is RG flow where the bulk flows become massless integrable flow\cite{Zamolodchikov:1989hfa}. As analytical methods, the importance of thermodynamic Bethe ansatz \cite{Yang:1968rm,Bajnok:2010ke,vanTongeren:2016hhc} and truncated conformal space approach \cite{Yurov:1989yu,Yurov:1991my,James_2018} should be noted. In the following subsections, we revisit several examples in contemporary condensed matter physics and related fields. One can see simplification of the discussions benefitted from our formalism.

\subsection{Edge modes of symmetry enriched criticality}
In \cite{Verresen:2019igf}, a (symmetry) protected edge mode for a quantum spin chain described by the Ising conformal field theory has been proposed. In our formalism, the analysis of protected edge modes in their model is summarized as follows.

\begin{itemize}
\item{Choosing $|\alpha_{i}\rangle= |I\rangle + | \psi\rangle$, where I and $\psi$ correspond to the primary fields of $M(3,4)$ CFT with conformal dimension $h_{I}=0$ and $h_{\psi}=1/2$. }
\item{Restricting the form of boundary operator by assigning nontrivial symmetry induced from lattice model. In their case, the boundary term has been restricted by the usual $Z_{2}$ spin flip symmetry and the complex conjugate symmetry.}
\end{itemize}
By the second operation, the boundary order and disorder operator with conformal dimensions $1/16$, and $1/2$ are excluded. Hence, the boundary-relevant operator can be restricted by this nontrivial symmetry, and one can apply the finite size scaling analysis to this non-Cardy state. Surprisingly, they have excluded the irrelevant boundary perturbations up to the seventh-level descendant operator. 

However, it should be mentioned that this kind of non-Cardy state can exist ubiquitously in systems with non-invertible topological defects (defect analog of non-abelian anyon, roughly speaking), and one can protect them by assigning symmetry act on the boundaries commonly. One can see related proposals in \cite{Yu:2021rng,Wu:2023ezm,Huang:2023hqx} and the appearance of such states under boundary RG in general\cite{Graham:2001pp,Recknagel:2000ri}. Moreover, in the Ising model, one can prohibit the boundary-relevant terms only assuming usual $Z_{2}$ and Kramers-Wannier $Z_{2}$ symmetry without introducing the complex conjugate symmetry\cite{Fukusumi:2020irh}.

\subsection{Bulk induced boundary perturbations and avoidance of Van Kampen's obstraction}

Recently, the difficulties of applying Kubo's response theory \cite{doi:10.1143/JPSJ.12.570,10.1143/PTP.15.77,2020arXiv200310390W} to an interacting conductor have been revisited\cite{PhysRevB.104.205116,Liu:2021gyt,Fukusumi:2021qwa}. Historically, this problem was first proposed by Van Kampen \cite{Kampen_1971,VANVELSEN1978135}. To some extent, topological invariants in modern condensed matter physics, such as the TKNN number\cite{Thouless:1982zz}, have been introduced to avoid this problem \cite{1985AnPhy.160..343K}.

The bulk RG understanding of this phenomenon is quite simple\cite{Fukusumi:2021qwa}. First, we consider a $1+1$ dimensional system with closed boundary conditions. Then, let us take the bulk perturbation as $H_{\text{B}_{UV}}=H_{\text{B},mar}+H_{\text{B},irrev}$, where$H_{\text{B},mar}$ is a marginal perturbation corresponding to flux and $H_{\text{B},irrev}$ is an irrelevant term corresponding to the interaction. In the response theory, one has to treat $H_{\text{B},irrev}$ first before the marginal term $H_{\text{B},mar}$. So this formulation is inapplicable to a large system with finite flux analyzed by the Bethe ansatz\cite{ALCARAZ1988280}, which is a central concern of condensed matter physicists (Readers interested in the historical aspect of this puzzle can see the discussions in \cite{Fukusumi:2021qwa}).

However, it is known that one can avoid this difficulty by considering a system with open boundaries or a repulsive defect\cite{2000LNP...544....3S,takasan2021adiabatic}. This can be understood as the bulk-induced boundary RG studied in $D$-brane\cite{2006hep.th....9034F,2015JHEP...12..114K}. The following bulk and boundary RG gives a concise understanding,

\begin{align}
&\mathcal{H}_{\text{CFT},\alpha_{0},\alpha_{1}}\rightarrow \mathcal{H}_{\text{CFT},\alpha'_{0},\alpha'_{1}} \\
\begin{split}
&H_{\text{CFT},\alpha_{0},\alpha_{1}}+H_{\text{B}_{UV}}+H_{\text{b}_{UV},0} +H_{\text{b}_{UV},1}\\
&\rightarrow H_{\text{CFT},\alpha'_{0},\alpha'_{1}}+H'_{\text{B}_{IR}}+H'_{\text{b}_{IR},0} +H'_{\text{b}_{IR},1}
\end{split}
\end{align}
where all perturbations labeled by IR are irrelevant. In this process, it is remarkable that the effect of the bulk marginal perturbation results in the change of boundary perturbations.

\subsection{Edge modes in topologically ordered systems and entanglement surface of anomalous theory}

The Majorana or parafermion edge modes in one-dimensional quantum systems have captured attention in theoretical condensed matter, because of their topological protectedness. In this type of model, one can choose the boundary state of UV CFT as,
\begin{equation}
|\alpha_{i}\rangle =\sum_{p=0}^{N-1}(D_{J})^{p}|I\rangle,
\label{UV_edge_modes}
\end{equation}
where $D_{J}$ is the $Z_{N}$ topological defect labeled by the $Z_{N}$ symmetry generator $J$ of the model and $|I\rangle$ is the Cardy's boundary condition corresponding to the identity operator (more generally, one can replace $I$ to primary fields with $Z_{N}$ charge zero). Without (generalized) symmetry at the boundary, this boundary condition itself is unstable at criticality but stabilized under the bulk RG flow. When restricting our attention to the boundary, the $Z_{N}$ symmetry preserved RG indicates the existence of the following bulk and boundary RG flow,
\begin{equation}
|\alpha_{i}\rangle\rightarrow |\alpha'_{i}\rangle =\sum_{p=0}^{N-1}(D_{J})^{p}|I'\rangle,
\label{edge_modes}
\end{equation}
where $I'$ reflects the flow of identity operator $I$ at UV.

By taking the parafermion basis in the quantum lattice model, one can observe its topological protectedness whereas it can be interpreted as spontaneous symmetry breaking in spin chain basis. It is worth noting that the structure $\sum_{p=0}^{N-1}(D_{J})^{p}$ plays a central role in considering the Tambara-Yamgami category, and the corresponding boundary condition can naturally be interpreted as a $Z_{N}$ generalization of free boundary condition in the $Z_{N}$ simple current extended CFT\cite{Runkel:2020zgg,Fukusumi:2020irh,Fukusumi_2022_c}.

Moreover, one can observe the absence of $Z_{N}$ invariant Cardy's states in anomalous theory, and the possible $Z_{N}$ invariant boundary states can be restricted in the form of Eq.\eqref{edge_modes} without extending the symmetry \cite{Han:2017hdv} (when extending or breaking the symmetry, one can see appearance of new boundary conditions \cite{Affleck1998,Iino:2020ipa} or symmetry breaking brane \cite{Quella:2002ns,Quella:2002fk,2002JHEP...06..028Q,Blakeley:2007gu} preserving $Z_{N}$ invariance). Hence the boundary states Eq.\eqref{edge_modes} appearing at the entanglement surface of topologically ordered state can be connected to the unstable boundary condition \eqref{UV_edge_modes} in bulk and boundary RG.

Similar analysis for nonanomalous theories (corresponding to CFTs with integer-spin simple currents\cite{Schellekens:1990ys,Gato-Rivera:1990lxi,Gato-Rivera:1991bcq,Gato-Rivera:1991bqv}) can be applied to the SPT phases, including gapless SPT\cite{Scaffidi:2017ppg}. For example, the double degeneracies in the entanglement spectrum of $S=1$ Haldane phase \cite{Pollmann_2010} can be connected to the boundary condition $|I\rangle +|\psi\rangle$ with $\sqrt{2}$ degeneracy for each edge\cite{Verresen:2019igf}. This state has a smaller boundary $g$-value compared with the boundary state $2|\sigma\rangle$ which corresponds to the edge spin $1/2$ degree of freedom appearing in the real boundary. Moreover, one can easily observe the existence of RG flow $2|\sigma\rangle\rightarrow |I\rangle + |\psi\rangle$ by applying the argument in \cite{Graham:2003nc}. Hence the edge modes appearing at the entangling surface can be more robust compared with those appearing at the real boundary by comparing their $g$-values. We expect that this analysis can be applied to SPT phases, including gapless SPT phases in general.

It should be worth noting that the boundary states Eq. \eqref{edge_modes} can appear as bulk degenerate states under bulk RG flow, by applying the arguments by Cardy \cite{Cardy:2017ufe}. This observation has been first applied to a lattice model in \cite{Ares:2020uwy} as far as we know. Recent proposal based on Cardy's arguments and anomaly analysis \cite{Furuya:2015coa,Numasawa:2017crf,Yao:2018kel,Kikuchi:2019ytf} can be seen in \cite{Li:2022drc,Kikuchi:2022ipr,Fukusumi_2022_c}.

\section{Modern understanding of duality}
\label{section_duality}

In this section, we study the implication of dualities in our formalism. For a system with open boundary conditions, it has been proposed that the duality transformation can be implemented by the shift of topological defects from one boundary to the other. This interpretation of dualities has appeared in \cite{Frohlich:2004ef,2007NuPhB.763..354F}, and applied to one-dimensional systems in \cite{Fukusumi:2020irh,Li:2023mmw,Li:2023knf}. We also note several related works that have studied the application of topological defects to bulk fields and Hilbert space and their realization in the lattice models \cite{Buican:2017rxc,PhysRevB.94.115125}. We concentrate on the defect realization of topological symmetry line or generalized symmetry which extends to the time direction\cite{https://doi.org/10.48550/arxiv.2003.11293,Belletete:2018eua}. This application of topological defects may not be a standard in the fields of high-energy physics and theoretical condensed matter interested in the generalized symmetry (or the nonlocal integral of motions), but it is relevant to implement duality for the system with open boundary conditions. (For the readers interested in the realization of topological symmetries as the integral of motions, we note recent work analyzing bond algebra\cite{Lootens:2021tet,Lootens:2022avn}.) 

Let us assume the following protected bulk and boundary RG from UV CFT to IR $\text{QFT}_{\mathbf{A}}$,

\begin{align}
&\mathcal{H}_{\text{CFT}_{UV},D_{\sigma}\alpha_{0},\alpha_{1}}\rightarrow \mathcal{H}_{\text{QFT}_{A},(D_{\sigma}\alpha_{0})',\alpha'_{1}} \\
\begin{split}
&H_{\text{CFT}_{UV},D_{\sigma}\alpha_{0},\alpha_{1}}+H_{\text{B}_{UV}}+D_{\sigma}H_{\text{b}_{UV},0} +H_{\text{b}_{UV},1}\\
&\rightarrow H_{\text{QFT}_{\mathbf{A}},(D_{\sigma}\alpha_{0})',\alpha'_{1}}+H'_{\text{B}_{\mathbf{A}}}+H'_{\text{b}_{\mathbf{A}},0} +H'_{\text{b}_{\mathbf{A}},1} 
\end{split}
\end{align}
where we have taken the left boundary condition as $D_{\sigma}\alpha_{0}$, which is constructed by applying topological defect $D_{\sigma}$ to the boundary constion $\alpha_{0}$. The precise definition of $D_{\sigma}$ and its application to boundary operator can be seen in \cite{Graham:2003nc,Kojita:2016jwe}, but it should be stressed that their precise structure has never been studied systematically regardless of their importance to study or formulate protected edge modes. More historical aspects of this problem can be seen in the author's previous work\cite{Fukusumi:2020irh}.

Because of the transmissive property of the topological defect, one can move this defect from the left boundary to the right boundary at criticality (in the cases with lattice realizations, this shift of topological defect can be implemented by unitary transformations\cite{PhysRevB.94.115125,Cobanera:2009a,Cobanera:2011wn,Nussinov:2011mz,Cobanera:2012dc}. A notable example is Kennedy-Tasaki transformation\cite{Kennedy:1992ifl,1992CMaPh.147..431K}, which has been revisited in \cite{Li:2023mmw,Li:2023knf}). Hence by applying the technique in \cite{Graham:2003nc}, one observes the following protected bulk and boundary RG to the dual UV $\text{QFT}_{\mathbf{B}}$,

\begin{align}
&\mathcal{H}_{\text{CFT}_{UV},\alpha_{0},D_{\sigma}\alpha_{1}}\rightarrow \mathcal{H}_{\text{QFT}_{\mathbf{B}},\alpha_{0}'',(D_{\sigma}\alpha)''_{1}} \\
\begin{split}
&H_{\text{CFT}_{UV},\alpha_{0},\alpha_{1}}+D_{\sigma}H_{\text{B}_{UV}}+H_{\text{b}_{UV},0} +D_{\sigma}H_{\text{b}_{UV},1}\\
&\rightarrow H_{\text{QFT}_{\mathbf{B}},(D_{\sigma}\alpha_{0})'',\alpha''_{1}}+H''_{\text{B}_{\mathbf{B}}}+H''_{\text{b}_{\mathbf{B}},0} +H''_{\text{b}_{\mathbf{B}},1} 
\end{split}
\end{align}
where $D_{\sigma}H_{B_{\mathbf{B}}}$ is notifying the application of duality operation to bulk operators and we have labeled the dual IR theory $QFT_{\mathbf{B}}$ by using '' . In other words,
 we have clarified the duality of the RG as,
\begin{equation}
D_{\sigma} : \{\text{CFT}_{UV} \rightarrow \text{QFT}_{\mathbf{A}}\} \Rightarrow \{\text{CFT}_{UV} \rightarrow \text{QFT}_{\mathbf{B}}\}
\end{equation}
in an evident and treatable way.

In this formalism, one can relate the protected edge modes $(D_{\sigma}\alpha_{0})',\alpha_{1}'$ in a quantum phase to $\alpha_{0}'',(D_{\sigma}\alpha_{1})''$ in its dual quantum phase. This enables one to analyze protected edge modes in dual theories and clarifies the interpretation of topological defect (or topological symmetry line) as a duality operation.

\section{Toward higher dimensional system}
\label{section_higher_dimension}

\begin{figure}[htbp]
\begin{center}
\includegraphics[width=0.5\textwidth]{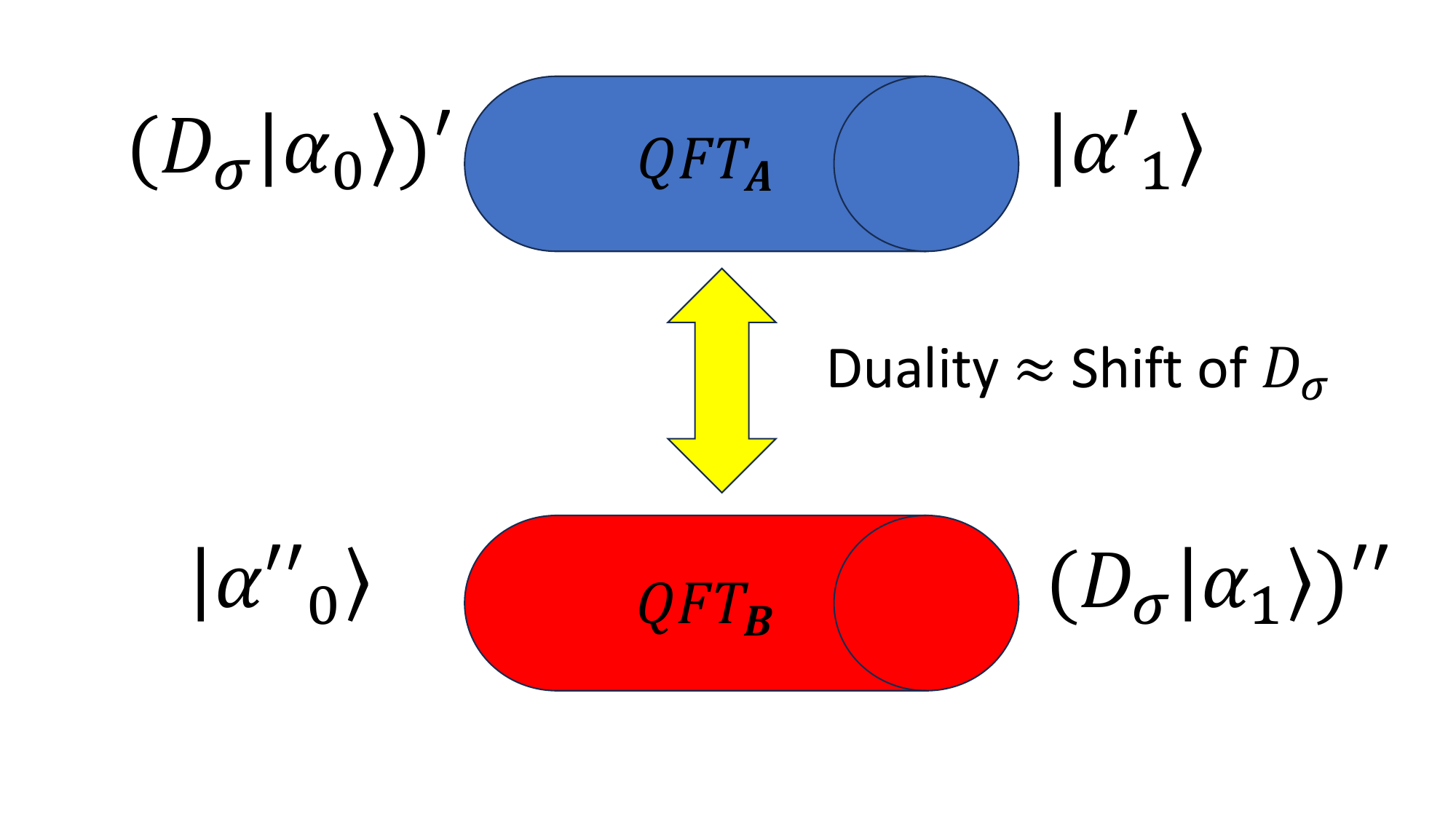}
\caption{An implementation of duality for general $D$ dimensional quantum field theory and its bulk and boundary RG flow. The boundary states live in $\mathcal{M}_{D-2}\times \{ 0\}$ and $\mathcal{M}_{D-2}\times \{ 1\}$  respectively and the $D-2$ dimensional defect which can be thought as a generator of duality moves from the left boundary to the right boundary.}
\label{Duality_picture}
\end{center}
\end{figure}

As a generalization, one can apply the arguments in the previous section to the system with higher space-time dimensions, by taking the space manifold as $\mathcal{M}_{D-2}\times [0,1]$, where $\mathcal{M}_{D-2}$ is a $D-2$ dimensional manifold. In this setting, the left and right boundary are $\mathcal{M}_{D-2}\times \{ 0\}$ and $\mathcal{M}_{D-2}\times \{ 1\}$  respectively (FIG. \ref{Duality_picture}). To implement duality in this system, it is necessary to introduce $D-2$ dimensional transmissive objects that can act on the  $\mathcal{M}_{D-2}\times \mathcal{T}$ where we have taken time direction as torus to consider quantum Hamiltonian formalism (When considering the Lagrangian formalism, this transmissive object lives in the manifold $\mathcal{M}_{D-2}\times \mathcal{T}$, by taking the time direction as torus $ \mathcal{T}$). Recent studies of generalized symmetry, such as categorical symmetry\cite{Kong:2014qka,Chatterjee:2022tyg,Chatterjee:2022kxb,Zeev:2022cnv} and noninvertible symmetry can be a good starting point for the application of our formulation. There exist many works studying them so the readers can see the reviews \cite{McGreevy:2022oyu, Cordova:2022ruw} and reference therein (For the readers interested in their realization on the lattice models, we note \cite{Cobanera:2009a,Cobanera:2011wn,Nussinov:2011mz,Cobanera:2012dc}). In this setting, one can also insert objects with spacial dimension lower than $D-2$ at the boundary, and it might be interesting to consider the effects of the shift of such objects at both bulk and boundary.

Here we comment on the correspondence between the RG flow of CFT and BCFT which has been proposed by Cardy\cite{Cardy:2017ufe}. Cardy's correspondence results in a nontrivial correspondence between $BCFT_{D}$ and its RG flow. In other words, for the bulk part, one can expect the following relations,
\begin{equation}
|B\rangle \sim QFT_{D}
\end{equation}
where $|B\rangle$ is the boundary state which corresponds to the bulk RG flow.

Hence by using the same notations in Sec.\ref{section_duality}, we can obtain the following correspondence between the theory $QFT_{\mathbf{A}}$ and $QFT_{\mathbf{B}}$ with duality,

\begin{align}
D_{\sigma}: \{ (D_{\sigma}|\alpha_{0}\rangle)' ,|B_{\mathbf{A}}\rangle ,|\alpha'_{0}\rangle \} \Rightarrow \{ |\alpha''_{0}\rangle ,|B_{\mathbf{B}}\rangle ,(D_{\sigma}|\alpha_{1}\rangle)'' \} 
\end{align}
where we have labeled the boundary states related to the bulk RGs by the corresponding QFTs. (Naively, we expect $D_{\sigma}|B_{\mathbf{A}}\rangle=|B_{\mathbf{B}}\rangle$ or $D_{\sigma}|B_{\mathbf{B}}\rangle=|B_{\mathbf{A}}\rangle$, but this needs further investigations.)

When considering the flow between $CFT_{D}$ to $TQFT_{D}$ with $CFT_{D-1}/TQFT_{D}$ correspondence\cite{Witten:1988hf}, one can see the bulk and boundary RG flow as,
\begin{equation}
|\alpha'\rangle \sim \mathcal{H}_{CFT_{D-1}}.
\end{equation}
when restricting our attention to one boundary.
In this way, one can connect boundary state $|\alpha\rangle$ of $BCFT_{D}$ to the Hilbert space of $CFT_{D-1}$ under the RG flow. This can be thought of as a generalized argument in \cite{Li:2022drc}. This may give a criterion to analyze general non-Cardy boundary states, which are out of the scope of the present BCFT analysis.

 Moreover, if two RG flows related by the duality result in different $TQFT_{D}$, this implies a correspondence between two different $CFT_{D-1}$ at the boundaries. In other words, this observation may give some clues to develop an understanding of the nontrivial (or mysterious) correspondence between general CFTs with different central charges. For example, we can list the level-rank duality of $SU(N)_{K}$ WZW model and $SU(K)_{N}$ WZW model\cite{Naculich:1990hg,Nakanishi:1990hj}, correspondence between the symplectic fermion and Dirac fermion\cite{Guruswamy:1996rk}, and construction of unitary CFTs from non-unitary CFTs by using the Hecke operation \cite{Harvey:2019qzs} (Galois symmetry and related ideas in the study of quantum groups may play a role in this kind of correspondence, but further systematic investigation may be necessary). For a more unified understanding of these correspondences, it may be important to analyze higher dimensional BCFTs with sufficient central charge and $g$-value and their bulk and boundary RGs. 

It is worth noting that the existence of a nontrivial relation between bulk TQFT and CFT at its boundary has been proposed in \cite{DiPietro:2019hqe,Behan:2020nsf,Behan:2021tcn}. Based on these works and the recent classification for the edge theory of TQFT \cite{Kaidi:2021gbs}, one can relate the puzzle of thermal Hall conductance in experimental settings in \cite{Mross_2018,Wang2017TopologicalOF} as this nontrivial relation between bulk and boundary theories\cite{Fukusumi_2022_c}. If this scenario is the case, this may imply the importance of the systematic RG analysis of the protected edge modes even in the experimental settings.

\section{Conclusion}
\label{section_conclusion}
In this work, we have proposed a formulation of protected edge modes by using bulk and boundary RG arguments. In our formulation, a series of nontrivial boundary phenomena in the contemporary condensed matter and related fields can be described in an evident and concise way. Hence it can be useful even for further numerical and experimental investigations. Moreover, our work gives a unified formulation and understanding of bulk and boundary RG phenomena in general space-time dimensions, in principle. Of course, one needs to modify or define our formulation more rigorously, corresponding to the specific settings or models, but we believe our formulation gives some clues to clarify the problems about protected edge modes in a tractable way. For this purpose, the construction of dualities by using bond algebra may be useful\cite{Cobanera:2011wn,Nussinov:2011mz,Cobanera:2012dc}.

Based on the modern view interpreting the RG as a projection of Hilbert space of UV theories, one can say the bulk and boundary symmetry analysis of higher dimensional conformal field theory can unify a series of dualities in contemporary mathematical physics and related fields. Hence we stress the necessity of further investigation of boundary conformal field theories in general space-time dimensions to understand systems at and away from criticality. Realizations of CFT in the lattice models in general space-time dimensions are limited \cite{Schuler:2016kuw,Thomson:2016ttt,Whitsitt:2017ocl,Zhu:2022gjc} and the application of the existing BCFT analysis \cite{Liendo:2012hy,Gliozzi:2015qsa,Herzog:2017xha,Bissi:2018mcq,Mazac:2018biw} to these models is challenging. To achieve their RG analysis it may be necessary to develop not only analytical methods but also numerical methods such as truncated conformal space approach\cite{Watts:2011cr,Hogervorst:2014rta} and tensor-network methods\cite{Iino:2019vxd,Chen:2022wvy,Yamada:2022dka} in higher dimensional systems with boundaries.

\section{acknowledgement}
We thank Linhao Li, Hosho Katsura, Masaki Oshikawa, and Yunqin Zheng for the helpful comments and discussions. We also thank Ji Guangyue, Bo Yang, and Shumpei Iino for the collaborations on past projects closely related to this project.

\appendix

\bibliography{bulk_boundary}

\end{document}